\documentstyle{ichep}
\input psfig
\def\sigtot{$\sigma_{\rm tot}$}
\def\sigtop{$\sigma_{t \overline{t}}$}
\def\pbarp{$\overline{p} p$}
\def\qqbar{$\overline{q} q$}
\def\ttbar{$t \overline{t}$}
\def\d0{D\O\ }
\def\ipb{pb$^{-1}$}
\def\met{$\not$E$_T$}
\def\pt{p$_T$}
\def\et{E$_T$}
\def\htran{H$_T$}
\def\gevcc{GeV/c$^2$}
\def\gevc{GeV/c}
\def\gev{GeV}
\def\njet{N$_{jet}$}
\def\apl{$\cal{A}$}
\begin{document}

\title{Search for the Top Quark: Results from the \d0 Experiment}

\author{P.D. Grannis}

\affil{State University of New York at Stony Brook}

\collab{For the \d0 Collaboration}

\abstract{We review the search for the top quark conducted by the
\d0 collaboration using data from the Fermilab \pbarp ~collider.  Based upon
a preliminary analysis of
an integrated luminosity of about 13.5 pb$^{-1}$, we have searched for
\ttbar ~production and decay in the experimental channels involving
a pair of dileptons ($e$ or $\mu$) plus jets, or single leptons plus jets.
Summed over all channels,
we observe 7 events in our data, to be compared with an
expectation from background processes of 3.2 $\pm$ 1.1 events.  The
\ttbar ~cross-section deduced from the small excess of events is presented
as a function of the top quark mass.  The statistics are sufficiently
limited that no clear evidence for the existence of the top quark can
be obtained.
\cabs
We also comment upon contributions to the Parallel session devoted to the top
quark at this conference.}

\twocolumn[\maketitle]

\section{Introduction}
We have a firm expectation that the top quark should exist; the Standard
Model (SM) requires it as the weak isospin partner for the $b$-quark,
completing the roster of three isodoublet quark and leptonic fundamental
fermions.  Extensions of the SM almost uniformly demand the existence of
the top as well.

We now strongly believe that the top is heavy.  The present experimental limit
\cite{dzerolimit} assuming SM production and decays is 131 \gevcc .
The CDF collaboration \cite{cdftop} has presented results which suggest
the possibility of top quark production with masses in the range
$160 \le {\rm m}_t < 190$ \gevcc .
Under the assumption that the  top is so massive,
certain simplifying features result.   The production of the top quarks
proceeds primarily through the pair production of \ttbar ,
initiated primarily by
\qqbar ~annihilation (and to some extent by gluon-gluon fusion processes)
\cite{laenen}.
Production of single top quarks through $W$-gluon fusion
\cite{singletop} is expected
to be small in comparison.
The decays of $t$ are simple in the SM: the $t$($\overline{t}$)
decays to $W^+ b$ ($W^- \overline{b}$) 100\% of the time, though new particles
outside the SM framework, such as
a charged Higgs boson
with mass below top mass m$_t$, could perturb the decay scheme.

At the partonic level, the final state reached after \ttbar ~decay
is controlled simply by the nature of the $W$ decays.  With both $W$'s decaying
leptonically, we expect two leptons, two jets due to the $b$'s, and missing
transverse energy (\met) due to the two neutrinos.  With one $W$ decaying
leptonically and the other hadronically, we expect a single lepton,
four jets and \met .  The channels in which both $W$'s decay hadronically
result
in six jets, but are experimentally difficult owing to the large multijet
production cross-sections.   In real life the simple partonic content in the
\ttbar ~decays can be modified by inefficiencies in jet reconstruction and by
the radiative emission of gluons from initial and final state partons.
The decay branching ratios are controlled by the branching ratio's
$W \rightarrow \ell \nu = 1/9$ for each lepton type, and
$W \rightarrow q \overline{q} = 2/3$ for the hadronic modes.  In the
experiments
only $e$ and $\mu$ decays of the $W$ are sought.

It is expected that for a top quark with mass above 130 \gevcc , the top
will decay so rapidly that the fragmentation of the top into hadrons
does not have time to occur \cite{panchieri}.  Thus the future study of top
quarks will afford a unique opportunity for investigation of bare quark states.

It has long been known that there remains a possibility for
${\rm m}_t<{\rm m}_W$
in the case that some unobserved particle exists into which
the top can decay \cite{hou}.  The total $W$ decay width
is sensitive to a contribution from $W \rightarrow t \overline b$; present
data from CDF and \d0 on $\Gamma_W$ \cite{widthw} limit possible
top masses to about $63 ~{\rm GeV/c}^2 \leq {\rm m}_t \leq {\rm m}_W$.

A large body of very precise data on electroweak processes has been assembled
over the past years which give rather stringent constraints upon the possible
values for m$_t$ -- {\it assuming the validity of the SM}.  The measurements of
the mass, width, and line shape of the $Z$ at LEP; asymmetries in the decay
distributions of fermions from the $Z$ at LEP and production asymmetries using
polarized electrons at SLC; the $W$ mass measurements from the Tevatron
collider
and CERN S$\overline{p}p$S; and neutrino scattering experiments are summarized
elsewhere in these Proceedings \cite{schaile}.
With the assumption that these phenomena
are correlated within the SM, one may infer the range
of possible top masses \cite{schaile}:
m$_t^{SM} = 178 \pm 11 ~_{-19}^{+18}$ \gevcc,
where the last error derives from the variation of the SM Higgs boson mass
between 60 and 1000 \gevcc .

Despite the indirect evidence for top it is necessary to pursue the direct
search with as little model prejudice as possible, since non-SM effects
could affect either the production or decay properties.
Indeed, with the expectation that the top is
heavy, the phase space available for new phenomena to alter either production
or decay schemes in enhanced.  In any case, the large Yukawa coupling expected
for a heavy top suggests that the top quark may play a special role in
the mass generation mechanism and therefore that its properties may be special.

\section{\d0 Search for the Top Quark}
The \d0 detector \cite{dzeronim}
is well suited for the search for the top quark.
The detector employs a finely segmented uranium-liquid argon calorimeter,
with uniform response to electromagnetic particles and hadrons for
$|\eta | < 4.2$.  The calorimeter
permits good multijet discrimination with relatively small corrections to the
observed jet energies.  Discrimination of electrons and pions is given by the
pattern of energy deposits in the calorimeter.  Muon candidates are confirmed
by their ionization in the calorimeter.  Good \met ~resolution
is achieved for signalling the presence of neutrinos, due to the good
energy resolution and hermetic calorimeter coverage.
The sagitta of muon trajectories is
measured using proportional drift
tube chambers before and after five magnetized iron toroids
surrounding the calorimeters.
The muon detector is sensitive over
the interval $|\eta| < 3.3$.  The large amount of material in the calorimeters
and toroids (between 13 and 18 absorption lengths) suppresses backgrounds
due to the leakage of hadronic showers.  The compact non-magnetic
tracking volume within the inner
calorimeter boundary is filled with drift chambers and a transition radiation
detector (TRD).  Its small outer radius helps to reduce the backgrounds
to muons from
$\pi$ and $K$ decays.  The tracking chambers serve to establish the primary
event vertex and confirm candidate lepton tracks.  The $dE/dx$ measurements in
the drift chambers and the signals from the TRD allow extra rejection of
background to electrons.

The search for the top quark reported here is based upon preliminary analyses
of data taken during the 1992-93 collider run.  We have optimized the selection
criteria for this search for top masses above 130 \gevcc ~in view of
the existing limit \cite{dzerolimit}.  The integral luminosity for these
searches is $13.5 \pm 1.6$ \ipb ~for the channels involving a $W$ decay
to electrons, and $9.8 \pm 1.2$ \ipb ~for those with $W \rightarrow \mu$.
The normalization of the luminosity scale has been set on the basis of
a weighted average of the available total cross-section measurements
at 1.8 TeV \cite{ezero}\cite{cdfsigtot}.  This average \sigtot ~is less than
the CDF measurement \cite{cdfsigtot} by about 6\%.  In calculating our \ttbar
{}~cross-sections we include a systematic error on luminosity of $\pm$12\%;
the effects of this luminosity error are small compared with our statistical
and
other systematic errors.

We report preliminary
measurements from three independent searches: the dilepton (all
combinations of $e$ and $\mu$) \cite{wimpenny}; the single lepton channel
(both $e$ and $\mu$) with topological cuts to suppress the background
\cite{serban}; and the single electron channel with $b$-jet tagging through
the semi-muonic decay of the $b$ \cite{raja}.

\subsection{Dilepton searches}

\begin{table*}
\vspace*{2pc}
\Table{|c|ccc|c|}{
 ~ & ~ & ~ & ~ & ~ \\
{\bf Mode} & $ee$ & $\mu\mu$ & $e\mu$ & All \\
 ~ & ~ & ~ & ~ & ~ \\
\hline
 ~ & ~ & ~ & ~ & ~ \\
{\bf Branching Ratio} & 1/81 & 1/81 & 2/81 & 4/81 \\
{\bf Acceptance} & 15\% & 9\% & 13\% & 13\% \\
No. \ttbar (m$_t$=160) & $0.22 \pm .04$ & $0.12 \pm .02$ & $0.40 \pm .05$ &
 $0.74 \pm .10$ \\
 ~ & ~ & ~ & ~ & ~ \\
\hline
 ~ & ~ & ~ & ~ & ~ \\
{\bf Main Bknds} & $Z \rightarrow \tau \tau$ & $Z \rightarrow \mu \mu$ &
$Z \rightarrow \tau \tau$ &  \\
  & Multijet fakes &   & $W^+ W^-$ & \\
 ~ & ~ & ~ & ~ & ~ \\
{\bf N$_{Bknd}$} & $0.16 \pm .07$ &  $0.33 \pm .06$ & $0.27 \pm .09$ &
 $0.76 \pm .13$ \\
 ~ & ~ & ~ & ~ & ~ \\
\hline
 ~ & ~ & ~ & ~ & ~ \\
{\bf DATA} & {\bf 0} & {\bf 0} & {\bf 1} & {\bf 1} \\
 ~ & ~ & ~ & ~ & ~ \\
}\caption{Expected top signal, backgrounds and observed events in the data
for the dilepton searches.\label{tabl1}}
\end{table*}

The dilepton mode analyses \cite{wimpenny}
require the presence of high \pt ~leptons,
large \met , and at least two jets with E$_T^{jet} > $15 \gev .  Additional
cuts are employed to suppress specific backgrounds arising from $Z$ decays,
cosmic rays and QED radiative processes.   Table 1 summarizes the dilepton
search results.

\begin{figure}
\vspace*{1pc}
\centerline{\psfig{figure=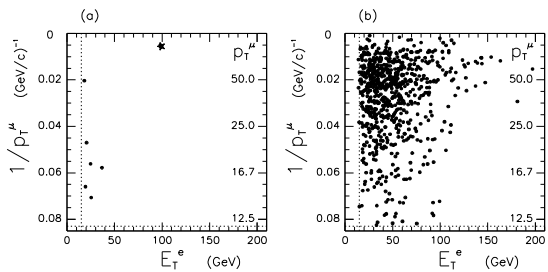,width=2.5in,height=1.1in}}
\caption{Distribution of events in E$_T^e$ and p$_T^\mu$ for the data
(before application of the final cut requiring two jets) and for Monte Carlo
(m$_t = 170$ $GeV/c^2$).  The Monte Carlo corresponds to about 1600 times the
luminosity for the data.\label{emu}}
\end{figure}

The data for the $e\mu$ mode before the final cut on $\geq 2$ jets,
and the expectation for m$_t = 170$ \gevcc ~taken from the ISAJET Monte Carlo
\cite{isajet}, are displayed in \Fref{emu} as a function of muon \pt ~and
electron \et .
The scale for the muon \pt ~is linear in (1/\pt ) since the measurement errors
are approximately constant and symmetric in this quantity.
Only the data event far from both electron and muon \pt ~cuts survives the jet
cut.  This event is quite striking; its kinematic parameters are
E$_T^e \approx 100$ \gev , p$_T^\mu \approx 200$ \gevc , \met $\approx 120$
\gev , and two jets with \et ~= 25 and 22 \gev .
Although the lepton transverse
momenta for this event are quite large, the overall likelihood for this event
agrees well with the expectations for SM top production,
considering the values of all 14
kinematic variables \cite{dzerolimit}.
The backgrounds from $Z \rightarrow \tau \tau$ are ruled
out for this event due to the large invariant mass of the $e \mu$ pair.
Over the full kinematic range for the selection of events in
this mode, we calculate that the ratio of expected number of
\ttbar ~(m$_t = 160$ \gevcc ) events to the number of backgrounds is
$4 \pm 2$.  Doubling the kinematic cuts on electron, muon and missing \et
{}~leaves the candidate event still far from the cuts and yields a
top/background ratio of $16 \pm 6$.  Interpreting this event as SM
top and performing a likelihood analysis for m$_t$ \cite{dzerolimit} ~gives
a large central value for the mass
(in the vicinity of 150 \gevcc ) with a relatively broad dispersion.

\subsection{Lepton + jets searches}
For the modes in which one $W$ decays leptonically and the other hadronically,
the branching ratio is greater than for the dilepton modes (12/81 for each
lepton type) but the backgrounds are larger.  There are two primary
background sources.
The first is due to the QCD (Drell-Yan) production of a $W$ in
association with the requisite ($\sim 4$) number of jets.  This process gives
exactly the same final state objects as the \ttbar ~signal, though the heavy
quark content and topology may be different.  The second is due to QCD
production of multijets (\njet ~$\sim 5$) in which one of the jets is
misidentified as a lepton and instrumental effects simulate sufficient
\met ~to satisfy the neutrino requirement.   The latter background afflicts
primarily the searches in the electron final states.
\d0 has performed two independent searches in the lepton + jets mode.  The
first
employs the differences in event topology to suppress backgrounds and tag the
top signal events.  The second uses tagging of $b$-quark jets through their
semileptonic decays into muons to reject the QCD backgrounds, which are
expected
to be less rich in heavy quark content.

\subsubsection{Topological tagging for lepton + jets. ~~~~~~~~~~~}

\begin{figure}
\centerline{\psfig{figure=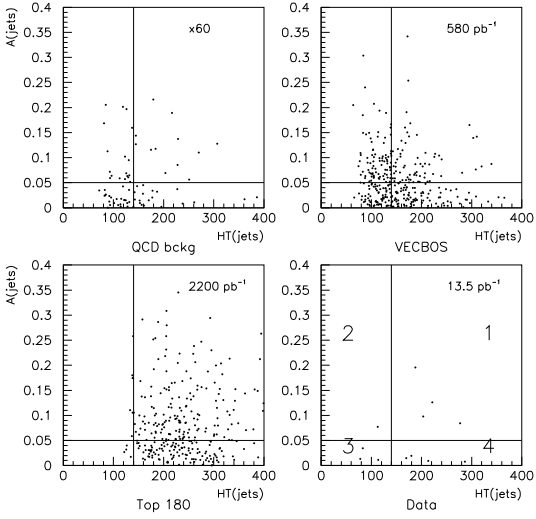,width=3.0in,height=3.0in}}
\caption{Event distributions {\it vs.} \apl ~and \htran ~for QCD multijets
(upper left), $W$+jets (upper right), a \ttbar ~Monte Carlo sample (lower left)
and for data (lower right).\label{aht}}
\end{figure}

The topological tag search \cite{serban} selects $W +$ jets final state
(both $e$ and $\mu$ decays) with the following criteria:
{\it (a)} large lepton momenta ~(E$_T^e > 20$ \gev ~and
$|\eta_e| < 2.0$ or p$_T^\mu > 15$ \gevc ~and $|\eta_\mu | < 1.7$);
{\it (b)} large \met ($ > 25$ \gev ~ for the electronic mode and
          $ > 20$ \gev ~ for the muonic mode); and
{\it (c)} at least 4 jets with \et ~$> 15$ \gev ~and $|\eta_{jet}| < 2.0$.
The {\it absence} of a $b$-tag is
required (see below) to preserve the independence of
this search from the $b$-tag analysis.   Finally, two variables describing the
topology of the event are defined.   The aplanarity (\apl ),
introduced for the study of
event shapes in $e^+ e^-$ experiments, is defined as 1.5 times the smallest
normalized eigenvalue of the momentum tensor,
constructed in the overall \pbarp ~frame  from the observed
jets with $|\eta | < 2$ in the event.
The cut chosen is \apl ~$> 0.05$.  The \ttbar ~events
tend to be more spherical (larger \apl ) than
the backgrounds which derive from QCD radiative processes which show
more tendency towards collinearity.
The variable \htran ~is defined
as the sum of the scalar transverse momenta of all
final state jets observed for $|\eta | < 2$
in the event.  Large \htran ~is indicative
of the decay
of a high mass state, and thus favors the \ttbar ~process.  The cut is chosen
at \htran ~$> 140$ \gev .

Distributions of events in the \apl -\htran ~plane are shown in \Fref{aht}
for Monte Carlo simulation of the QCD multijet background, the $W$+jets
process, and for \ttbar ~(m$_t = 180$ \gevcc ) production.  The \d0 data
distribution is also shown.  The cuts on \apl ~and \htran ~are shown; the
signal
region is above and to the right of the lines.

Two nearly independent methods have been used to estimate the backgrounds for
the topological lepton plus jets search.  The first proceeds from the
observation \cite{berends} that for the $W$+jets processes, the reduction in
cross-section upon requiring an additional jet is the same, independent of the
number of jets.  This is a natural consequence of QCD radiative processes,
since qualitatively, the
emission of each added jet incurs an additional factor of
$\alpha_s(q^2)$ ($q^2 \sim ({\rm m}_W)^2$).
Although the appropriate $q^2$ may vary somewhat with
\njet , this scaling law is found to be satisfied theoretically to within
20\% for up to four jets.   We refer to this behavior as `jet-scaling'.

\begin{figure}
\centerline{\psfig{figure=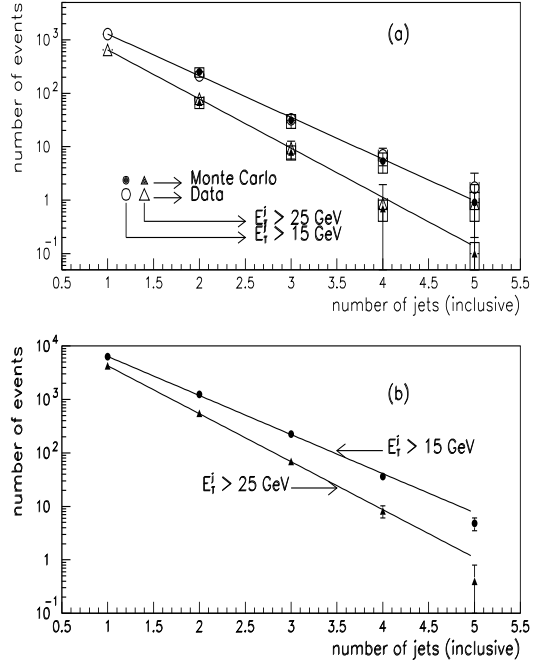,width=3.0in,height=4.0in}}
\caption{(a) Number of $W$+jets events (with $W \rightarrow e \nu$)
{\it vs.} the inclusive number of jets for
\et $>$ 15 \gev ~(upper points) and \et $>$ 25 \gev ~(lower points).
The open symbols denote data and the filled symbols show the prediction
of the Monte Carlo.  The lines are fits to the data for the interval
$1 < {\rm N}_{jet} < 3$.
(b) Number of multijet events {\it vs.} the inclusive number of jets for
\et $>$ 15 \gev ~(upper points) and \et $>$ 25 \gev ~(lower points).
The filled symbols denote data and lines are fits to the data for the interval
$1 < {\rm N}_{jet} < 4$.\label{jetscale}}
\end{figure}

In \Fref{jetscale}(a) we show the $W$+jets data (for $W \rightarrow e \nu$)
before application
of the \apl ~and \htran ~cuts, as a function of the
`inclusive' jet multiplicity (we plot the number of events with
\njet $\geq n$ at abcissa $n$) for two different jet \et ~thresholds.
The open symbols represent the data.  The filled symbols give the
predictions from the Monte Carlo, for which
we use the tree-level parton generator
VECBOS \cite{vecbos} with subsequent parton showering and
fragmentation of the partons
in ISAJET and passage of the resulting particles through a GEANT-based
\cite{geant} simulation of the detector.   The lines are fits to the
data for ($1 \leq {\rm N}_{jet} \leq 3$) and show good agreement with the
jet-scaling hypothesis for \njet $\leq 3$.   The extrapolation to \njet
{}~$\geq 4$ shows that the excess of events to be attributed to \ttbar ~in
this data selection is not large.   It is noteworthy that the data and Monte
Carlo predictions agree very well, both in absolute normalization and
the slope.

One may expect that similar jet-scaling behavior would arise for the QCD
production of multijets.  In \Fref{jetscale}(b)
we show the dependence on inclusive
jet multiplicity for a sample of multijet events in which one of the jets
has fluctuated to resemble (with bad $\chi^2$) an electron and for which
\met $< 25$ \gev .   For this process, which should contain no anomalous
signals at large \njet , the jet-scaling hypothesis again works well and the
slope of the distribution is very similar to that for the $W$+jets process.
Similar behavior is observed with poorer statistics
in the jet-scaling of a $Z$+jets sample where no
\ttbar ~production is expected.

The jet-scaling estimate of the background proceeds from the assumption that
the
$W$+jets events (and residual QCD multijet contributions)
satisfy the jet-scaling hypothesis, while the \ttbar ~events
contribute to a given multiplicity \njet ~according to fractions determined
from
Monte Carlo calculations.   The calculation yields the number of background
and \ttbar ~signal events surviving in the lepton + 4 or more jet sample.
The resulting
background estimate is then corrected for the probability that the
background events survive the \apl ~and \htran ~cuts (taken from Monte Carlo).
The resulting background estimated for the topological tagged experiment
is $1.8 \pm 0.8 \pm 0.4$; the systematic error includes the uncertainty in the
jet-scaling hypothesis.

The second method for estimating the background for the topological tag
is independent of the jet-scaling hypothesis.  From the Monte Carlo
distributions of events from QCD multijets, $W$+jets and \ttbar ~production
shown in \Fref{aht}, we deduce the {\it fraction} of events in each process
which fall into each of the quadrants 1 -- 4 of \apl -\htran ~space shown
in \Fref{aht}.  We then fit the data distribution in \apl -\htran ~space
(also shown in \Fref{aht}) using these fractions with the
background and \ttbar ~populations as free parameters.  We obtain in this way
an estimated background of $1.7 \pm 0.8 \pm 0.4$ events.  This second
estimate agrees very well with that obtained from jet-scaling above.

We observe a total of four events in our topological tag analysis (two
are $e$+jets+\met ~and two are $\mu$+jets+\met ).

\subsubsection{$b$-quark tagging for lepton + jets. ~~~~~}

Since \ttbar ~events are expected to be enriched in $b$-quarks relative
to the backgrounds, effective reduction of the backgrounds can be achieved
by tagging the presence of $b$'s in the $W$+jet sample.
The current \d0 $b$-tagging analysis \cite{raja} is performed
for the $W \rightarrow e$+jets sample using the inclusive
semileptonic decay  $b \rightarrow \mu$.

The event selection retains the large \et ~cut ($> 20$ \gev ) for the electron
but relaxes the \met ~cut to \met $> 20$ \gev ( $>35$ \gev ~if
$\phi_{\mu\nu}< 25^\circ$).   Only three jets are required with a threshold
\et $> 20$ \gev .   The topological cuts on \apl ~and \htran ~are not
required.   A muon consistent with the expectation for
$t \rightarrow b \rightarrow \mu$ is required in the event: p$_T^\mu > 4$
\gevc ~with $|\eta_\mu|<1.7$.   In the case that p$_T^\mu > 12$ \gevc ,
we require a separation between the muon and jet to be less than 0.4
in $\eta$--$\phi$ space to keep this analysis independent of the
$e\mu$ dilepton search.

The $b$ tagging rate is studied for several data sets involving multiple jets.
The tag rate is measured using multijet data in our $e$+jets trigger sample,
for which the electron is classified as fake (this sample is almost purely due
to multijet background).  The probability (per jet) for finding a tagging muon
is established from these data; it depends upon the \met ~cut imposed,
and upon the \et ~of the jet.   Comparison of this tagging rate for
flavor-undifferentiated processes can be made with independent data
samples of di-jet triggers, $\gamma$+jets and $Z$+jets.  The
VECBOS/ISAJET/GEANT Monte Carlo chain can be used to calculate the tagging
probabilities for the $W$+jets process.

\begin{figure}
\vspace*{5pc}
\centerline{\psfig{figure=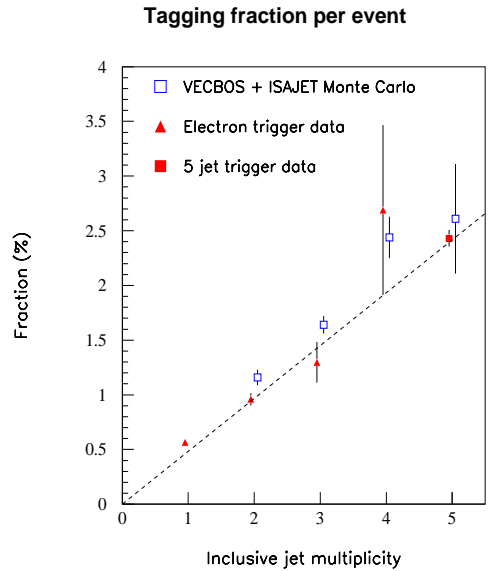,width=3.0in,height=2.5in}}
\caption{The probability per jet to observe a $\mu$-tag for a sample of fake
electron + jets events (due to QCD multijet production) (triangles);
QCD multijets from a five jet trigger (closed squares); and from the VECBOS
Monte Carlo simulation of the $W$+jets process.\label{mutag}}
\end{figure}

These tagging probabilities are shown in \Fref{mutag}.  The rates observed for
the different data samples agree and are also in good agreement with the Monte
Carlo expectations.   The tagging rates are about 0.5\% per jet and are in
agreement with the hypothesis that most of the muons come from $b$ decay.
There is no discernible difference in the tagging rates for QCD multijets
and $W$+jets.

Using ISAJET  and our detector simulation to determine the tagging rates
for \ttbar ~production, we find that the tagging rate
(per \ttbar $\rightarrow e$ + \met ~+ jets event) is about
20\%; it varies somewhat with the mass of the top quark due to the dependence
of
tagging probabilities with jet \et .   We note that since there are two
$b$'s per \ttbar ~event, and each $b$ can decay directly to muons or
via the cascade $b \rightarrow c \rightarrow \mu$ chain, the tagging rate
before
experimental selection cuts and inefficiencies would be expected to be over
40\%.   A confirmation of our $\mu$-tagging calculation can be found in the
determination of the inclusive $b$-quark production cross-section
\cite{hedin}, in which the tagging techniques are similar.   In this
separate study, the distributions as a function of \pt ~and p$_T^\mu$ relative
to a nearby jet are shown to conform to the expected mix of subprocesses,
and the resulting $b$ cross-section is in good agreement with NLO QCD
predictions \cite{nde}.

The backgrounds from multijets and $W$+jets in the $b$-tagging analysis
are determined by first separating the two background subprocesses
in the data sample, based
on their different \met ~distributions, and then applying the
relevent $\mu$-tagging probabilities (discussed above) as derived from data.
We find that the multijet background contributes $0.12 \pm 0.05$ events
and the $W$+jets background gives $0.43 \pm 0.14$ events.  The total
background of $0.55 \pm 0.15$ events is to be compared with the two events
observed in the \d0 data.

A cross check of the background estimates can be made by estimating the
background for the \njet ~$\geq 1$ and \njet ~$\geq 2$ samples where little
\ttbar ~contamination should be present.  Within errors, these estimates
agree with the data.  When these data for \njet ~$\leq 2$ are extrapolated to
\njet ~$\geq 3$  using the jet-scaling hypothesis, we obtain an estimated
background in excellent agreement with that deduced above.

\subsection{Cross-section results}

The summary of expected \ttbar ~signal events, background estimates, and events
observed in the \d0 data sample are summarized in Table 2.  We observe seven
signal events in the three independent analyses and expect a total of
$3.2 \pm 1.1$ background events in this preliminary analysis.
The significance of the excess 3.8 events over
background can be assessed by calculating the probability that our expected
background fluctuates to give at least seven data events, taking into account
the Gaussian errors on backgrounds, acceptances and luminosity, and the Poisson
errors on the number of events.  This probability is 7.2\% and corresponds
to about 1.5 standard deviations in the Gaussian approximation.  We conclude
that the \d0 experiment does not give significant evidence for an excess of
events to be attributed to \ttbar ~production.

\begin{table*}
\Table{|c|c|c|c|c|}{
 ~ & ~ & ~ & ~ & ~ \\
{\bf Analysis} & {\bf Dilepton} & {\bf $\ell$ + jets} &
                                        {\bf $\ell$ + jets} & {\bf All} \\
{}~~~ & {\bf Search} & {\bf Topological} & {\bf $b$-Tag} & {\bf Searches} \\
 ~ & ~ & ~ & ~ & ~ \\
\hline
 ~~ & ~~ & ~~ & ~~ & ~~ \\
{\bf N}(m$_t = 140$) & 1.4 $\pm$ 0.2 & 4.2 $\pm$ 0.9 & 1.3 $\pm$ 0.4 &
       6.7 $\pm$ 1.2 \\
{\bf N}(m$_t = 160$) & 0.7 $\pm$ 0.1 & 2.8 $\pm$ 0.5 & 1.0 $\pm$ 0.2 &
       4.4 $\pm$ 0.7 \\
{\bf N}(m$_t = 180$) & 0.4 $\pm$ 0.1 & 1.5 $\pm$ 0.3 & 0.6 $\pm$ 0.2 &
       2.5 $\pm$ 0.4 \\
 ~~ & ~~ & ~~ & ~~ & ~~ \\
\hline
 ~~ & ~~ & ~~ & ~~ & ~~ \\
{\bf Background} & 0.8 $\pm$ 0.1 & 1.8 $\pm$ 0.9 & 0.6 $\pm$ 0.2 &
3.2 $\pm$ 1.1 \\
 ~~ & ~~ & ~~ & ~~ & ~~ \\
\hline
 ~~ & ~~ & ~~ & ~~ & ~~ \\
{\bf DATA} & {\bf 1} & {\bf 4} & {\bf 2} & {\bf 7} \\
 ~~ & ~~ & ~~ & ~~ & ~~ \\
}\caption{Expected top signal for m$_t$ = 140, 160, 180 \gevcc ; expected
backgrounds, and observed events for each of the three \d0 top search
analyses.\label{tabl2}}
\end{table*}

\begin{figure}
\vspace*{5pc}
\centerline{\psfig{figure=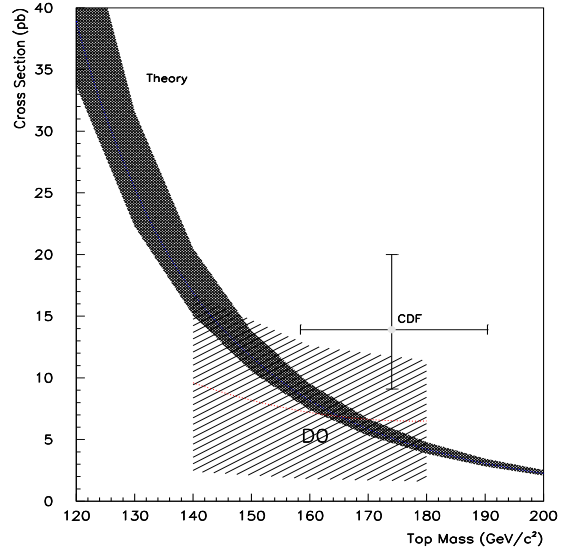,width=3.0in,height=2.5in}}
\caption{Cross-section {\it vs.} m$_t$.
The dotted line and cross-hatched area give the
\d0 preliminary result for a range of possible top quark masses.
The shaded band represents the
theoretical NLO calculation of Ref [3] with its estimated
theoretical errors.
The CDF cross-section and mass value [2] are shown as the data point with
errors.\label{topxs}}
\end{figure}

We can transform our counting results above into a cross-section
for \ttbar ~production by dividing the background-subtracted number of
events by the integrated luminosities and acceptances determined from
the ISAJET Monte Carlo \cite{marchesini}.
The acceptances vary with the assumed
top mass, so the cross-sections for the given number of events fall
somewhat as m$_t$ increases.  The preliminary results are shown in Table 3
and in \Fref{topxs}, together with the theoretical expectations \cite{laenen}
and the CDF result \cite{cdftop}.

We conclude from \Fref{topxs} that the preliminary \d0 data are consistent
with the SM expectations for m$_t$ above about
140 \gevcc , and can accomodate an
arbitrarily heavy top quark.   The \d0 result does not require
the invocation of new physics for \ttbar ~production.
The \d0 result is also consistent with the CDF
counting experiment  for \sigtop ~given the large errors on both
experiments.

It is of interest to compare the sensitivity for \ttbar ~production in \d0 with
that of CDF.
To do this, we can compare the expected number of \ttbar ~events
in the two experiments for a given choice of \sigtop .  For this purpose we
choose the CDF central value at m$_t$=174 \gevcc ~of \sigtop ~= 13.9 pb.
This result corresponds to the CDF observation of 12 candidate events.
After iterating the background estimate to allow for their excess signal
contributions, CDF estimates there are 3.4 background events and
8.6 \ttbar ~signal events.  With the same cross-section, \d0 would expect
3.2 background events and 8.2 \ttbar ~signal events.   Thus the two experiments
presently have comparable sensitivity.

\begin{table}
\Table{|c|c|}{
 ~~ & ~~ \\
 {\bf Top Mass} & {\bf Cross Section} \\
 {\bf (GeV/c$^2$)} & {\bf (pb)} \\
 ~~ & ~~ \\
\hline
 ~~ & ~~ \\
{\bf 140}  & {\bf 9.6 $\pm$ 7.2} \\
{\bf 160}  & {\bf 7.2 $\pm$ 5.4} \\
{\bf 180}  & {\bf 6.5 $\pm$ 4.9} \\
 ~~ & ~~ \\
}\caption{Preliminary results from \d0 for the \ttbar
{}~cross-section.\label{tabl3}}
\end{table}

{}From \Fref{topxs} it is apparent that the combined results of \d0 and CDF
for \sigtop ~give a result which is more consistent with the SM prediction
for \ttbar ~production with m$_t$ in the 140 - 180 \gevcc ~range than
the CDF results alone.  Performing an average of the two experiment values is
possible, but was not done at the time of the Conference.  It will
require a full-multichannel likelihood calculation for the several analyses of
both experiments, taking into account the effects of positive and negative
fluctuations of both signal and backgrounds.  It should be done using a proper
treatment of the correlated errors between the experiments and common choices
for background cross-sections and iteration procedures.   For these reasons,
and
because the \d0 results are preliminary at this stage, the average has not been
computed.   However, it seems likely that the use of the \d0 data will lower
the significance in the combined analysis compared to that from CDF alone.
I conclude that at this time, the experiments seeking the direct observation
of the top quark are not sufficiently sensitive to give solid direct evidence
for its existence.  It is however
true that the combined evidence from both experiments is
suggestive that the effects of \ttbar ~production are being observed in the
Fermilab Tevatron experiments.

\section{Comments on theoretical contributions to this Conference}
Stimulated by the CDF report \cite{cdftop} that the cross-section
for \ttbar ~production could be larger than expected within the SM
\cite{laenen}, several
suggestions have been made which invoke new physics, either raising
\sigtop ~or adding new processes which could mimic the top signature.   As
noted
above, we believe that in view of the \d0 results for \sigtop ~and the combined
errors in the experimental measurements and in the theoretical calculations, it
is by no means {\it necessary} at this time to invoke new physics.

Enhancements to the cross-section due to the production of new states
which decay into \ttbar ~pairs were discussed in two contributions to
this conference.  One \cite{lane} notes the possibility of producing the
`technieta' psuedoscalar particle ($\eta_T$),
required in Technicolor theories.  The second
invokes the possible existence of color-octet vector mesons (V$_8$)
\cite{parke}.   Both $\eta_T$ and V$_8$ decay dominantly into \ttbar ~pairs,
so would add to the experimentally observed cross-section.
In both cases, the production of the new objects in standard gluon fusion
and \qqbar ~annihilations can be computed; reasonable enhancements (of order of
a factor of 2) result for $\eta_T$ or V$_8$ masses in the vicinity of
500 \gevcc .

The possibility for new weak iso-singlet quarks (present in some
string-inspired
models) was noted \cite{barger}.   If these
exist, one would expect a full
set of all flavors of singlet quarks.  The production of singlet quarks
would be similar to the
``ordinary'' iso-doublet top quark  when their masses are near
m$_t$.   The decays of the iso-singlet quarks involve ordinary vector bosons
and
quarks.  For many flavors of the iso-singlet quarks,
the decay patterns would be expected to disagree with
known production and decay characteristics, but for the iso-singlet top in
particular it is possible to envision enhancement in the signatures expected
for the SM top quark.

Enhancement of ordinary top quark pairs would result in the case that anomalous
couplings are present between the \ttbar ~and scalar components of the theory.
Such anomalous couplings arising from dynamical symmetry breaking
considerations were examined \cite{arbuzov} and found to be capable of
producing
up to a factor of two increase in \sigtop .

Although we do not find that these mechanisms for signal enhancement are
presently warranted by the data, these interesting comments point to the rich
opportunities for the study of the top quark in the near future.   In
addition to increasing the signal cross-section,
specific new physics processes make
significant modifications to the top quark p$_T$ and angular distributions,
to the invariant mass distribution for \ttbar ~pairs, and can give interesting
departures from the SM decay patterns.   They thus serve to emphasize that the
direct searches for the top quark need to remain as free from
Standard Model bias as possible.   They reinforce the point that a massive
top quark, with possible decays into a variety of new objects ($e.g.$ the
charged Higgs) and its sensitivity to extra non-SM
production mechanisms,
is a fertile ground for direct observation of new physics.
The likely connection between a massive top quark and
the mechanisms of symmetry breaking also suggests that crucial new insights
could result from precision studies of its production and decay.   Finally,
the precision measurement of the mass of the top is crucial, since by
comparison
with the wealth of precision measurements in the electroweak sector one adds
powerful constraints on the validity of the SM and on the value of the Higgs
boson mass.

Prediction of the value of the top quark mass from general dynamical
arguments has by now a long history \cite{marciano}.  Two additional
predictions were presented to this conference.  The first
\cite{andrionov} exploits the likely
heaviness of the top quark and its large Yukawa coupling to the Higgs
to develop renormalization group constraints yielding
a top quark mass prediction of 170 $\pm$ 5 \gev .  The second
is based on a geometrical ``spin gauge'' model \cite{chisholm}
in which no Higgs bosons appear,
but sum rules involving fermions and gauge bosons can be derived.  In this
model the top quark mass is predicted to be $151.7 \pm 0.1$ \gev .   The
attempts
to calculate the top quark mass using dynamical or theoretical simplicity
arguments may ultimately help illuminate fundamental issues concerning
symmetry breaking in nature.  We hope that in the near future, the experiments
will have determined the mass with good precision and that
these theoretical issues can come into clearer focus.

Some phenomenological issues for the experimental
measurement of the mass of the top quark were discussed in this Conference.
As is well known, the simplicity of the \ttbar ~final
state at the partonic level ($e.g.$ for the lepton + jets mode,
a lepton pair from one $W$, a di-quark from the other $W$, and
a pair of $b$-quarks to be paired with each of the $W$'s) can be substantially
modified in the real experimental environment.   In addition to the detector
issues resulting from the definition of jets above a certain threshold in
\pt ~with sufficient isolation from other objects in the event, there are
complications from the possible presence of recognizable jets arising from QCD
radiation.   These gluon emissions can be from initial or final state
partons, or indeed, as stressed in this conference \cite{orr}, from
interference
diagrams which involve both initial and final state radiation.
The possible omission of some of the primary parton jets and
the possibile addition of radiative gluon jets compounds the combinatorial
problems of associating the right jets into $W$ and $t$ states.
Indeed Monte Carlo studies by the experimental groups have tended to show
that fewer than 50\% of the lepton + jets final states are reconstructed
correctly.   Improved techniques for identifying jets and their combinations
into parent objects will be most welcome when the top quark becomes solidly
established and attention turns to a precision measurement of its mass.

\section{Conclusions}
The search for the top quark in the \d0 detector has been made in three
independent decay modes:  dilepton decays of the \ttbar ~system;
single lepton decays with topological tags; and a $b$-tagging analysis
for the single electron decays.   The sensitivity of the experiment
is essentially the same as for the CDF analysis \cite{cdftop}.  The \d0
analysis
finds a total of 7 candidate events, to be compared with an expected background
of 3.2 $\pm$ 1.1 events.   The statistical significance of this result is not
sufficient to establish the top quark.   The \ttbar ~cross-sections
deduced from the analyses are consistent with the Standard Model predictions,
and with the CDF measurements.   Taking the two experiments together, we
find that there is not sufficient evidence to directly establish the existence
of the top quark, though the small excess of events are in reasonable agreement
with the indirect SM predictions based upon a variety of precision measurements
of electroweak parameters.

The \d0 results presented at this conference are preliminary.  Additional
analyses involving the muon + jets with a $b$-tag are in progress, as are more
sophisticated selections based upon multivariate analyses (Neural Networks,
Fisher discriminants, probability density estimators, etc.).  The analysis of
the lepton + jets events to yield mass estimates for a potential top quark are
in progress.  Most importantly, the Tevatron Collider is working efficiently in
the current experimental run expected to continue through mid-1995, and should
yield data sets of about four times that of the sample reported upon here.

\end{document}